\documentclass[aps,prb,twocolumn,floatfix,showpacs]{revtex4}
\usepackage{indentfirst}
\usepackage{epsfig}
\usepackage{amsmath}
\usepackage{graphicx}
\newcommand{\be}{\begin{equation}}
\begin{document}
\title{Doped carrier formulation of the t-J model:
the projection constraint and the effective Kondo-Heisenberg lattice
representation}
\author{Rafael T. Pepino}
\affiliation{International Center for Condensed Matter Physics,
Universidade de Brasilia, Caixa Postal 04667, 70910-900 Brasilia,
DF, Brazil}
\author{Alvaro Ferraz}
\affiliation{International Center for Condensed Matter Physics,
Universidade de Brasilia, Caixa Postal 04667, 70910-900 Brasilia,
DF, Brazil}
\author{Evgueny Kochetov}
\affiliation{Bogoliubov Theoretical Laboratory, Joint Institute for
Nuclear Research, 141980 Dubna, Russia\\ International Center for
Condensed Matter Physics, Universidade de Brasilia, Caixa Postal
04667, 70910-900 Brasilia, DF, Brazil}


\begin{abstract}
We show that the recently proposed doped carrier Hamiltonian
formulation of the t-J model should be complemented with the
constraint that projects out the unphysical states. With this new
important ingredient, the previously used and seemingly different
spin-fermion representations of the t-J model are shown to be gauge
related to each other. This new constraint can be treated in a
controlled way close to half-filling suggesting that the doped
carrier representation provides an appropriate theoretical framework
to address the t-J model in this region. This constraint also
suggests that the t-J model can be mapped onto a Kondo-Heisenberg
lattice model. Such a mapping highlights important physical
similarities between the quasi two-dimensional heavy fermions and
the high-T$_c$ superconductors. Finally we discuss the physical
implications of our model representation relating in particular the
small versus large Fermi surface crossover to the closure of the
lattice spin gap.
\end{abstract}
\maketitle

\section{Introduction}

The high-$T_c$ superconductors continue to be a puzzle to most
researchers in the field. Notably, in the underdoped regime, the
cuprates display highly anomalous physical properties. Above the
superconducting temperature, $T_c$, those lightly doped materials
are characterized by a spin gap and, by what seemed until recently,
disconnected Fermi arcs around preferential directions in momenta
space \cite{1,2}. This phase is widely referred to as the pseudogap
state. That such disconnected arcs are indeed integrated into one
coherent Fermi surface was recently demonstrated by the experiment
of Doiron-Leyraud et al \cite{3} unfolding the quantum oscillations
in the electrical resistance of YBa$_2$Cu$_3$O$_{6.5}$. The
pseudogap phase is in this way associated with a small Fermi surface
as opposed to the overdoped cuprates which exhibit large Fermi
surfaces instead \cite{4}. The evolution of the cuprate Fermi
surface as a function of doping has been monitored by several ARPES
experiments \cite{5,6}. Such a process has also been the object of
investigation in a number of renormalization group calculations
\cite{7}. Despite that, it is fair to say that a more qualitative
physical understanding of this Fermi surface crossover in the
cuprates is still lacking.

Ever since Anderson's suggestion that the high-$T_c$ cuprates are
driven by strong electron correlations \cite{8}, the Hubbard model,
or its large U version, the t-J model, have been the center of
attention of theorists. Recently, two new representations in terms
of dopant particles have been proposed
\cite{ribeiro05,ribeiro06,ferraz06} for the t-J model. In the
Hamiltonian approach put forward by Ribeiro and Wen
\cite{ribeiro05,ribeiro06}, the original projected electron
operators are replaced by spin-1/2 objects (called ``lattice
spins'') while the dopant particles are represented by fermions with
spin-1/2 (called ``dopons''). In the path-integral representation,
the spin-dopon partition function of the t-J model has been used to
formulate a resonating valence bond mean-field theory to describe
the superconducting phase in the cobaltates \cite{ferraz06}. The
motivation behind these approaches is clear: since the concentration
of dopons is small close to half-filling, the no double occupancy
(NDO) constraint for the dopons can be safely relaxed in that
regime.

However, in the description of electrons in terms of spins and
dopons, special care should be taken to avoid the inclusion of
unphysical states. As pointed out by Ferraz, Kochetov and Uchoa
\cite{ferraz07},  a constraint should be imposed to eliminate the
unphysical states. Within the doped carrier representation, the
conventional no double occupancy constraint for the lattice
electrons reasserts itself as the constraint to eliminate the
unphysical states in the enlarged spin-dopon Hilbert space.
Basically, it reflects the completeness relation of the physical
Hilbert space for the t-J model.

It is important to note that in the Hamiltonian doped carrier
representation of the t-J model the NDO constraint takes on a form
which is in a sense dual to that of the original one in the physical
lattice electron representation. The original local NDO constraint
for the physical lattice electrons,
$\sum_{\sigma}{c}_{i\sigma}^{\dagger}{c}_{i\sigma}\le 1$, cannot be
relaxed at half-filling (in fact, right at half-filling it reads
$\sum_{\sigma}{c}_{i\sigma}^{\dagger}{c}_{i\sigma}=1)$, but it can
be dropped safely at a large enough doping. In contrast, the new
spin-dopon constraint can be treated in a controlled way close to
half-filling (right at half-filling this constraint becomes a
trivial identity), whereas it cannot be safely relaxed in the
overdoped region. Since the underdoped region of the cuprate phase
diagram is of primary interest as was stressed earlier on, this new
doped carrier formulation of the t-J model, accompanied with the
spin-dopon constraint, sounds quite interesting and appealing.

Within the spin-dopon representation, the strong electron
correlation manifests itself in the constraint to exclude the
unphysical states. It turns out that this constraint can be treated
reliably at the mean-field level close to half-filing. This occurs
at the expense of having more complicated form of the t-J
Hamiltonian in this new representation.

The present paper demonstrates the significance of the constraint
that excludes the unphysical states in the doped carrier approach
from both the physical as well as the computational points of view.
Specifically, we show that there is a certain gauge ambiguity in
choosing a particular form of the unprojected t-J dopon-spin
Hamiltonian in the enlarged Hilbert space. This ambiguity is related
to a redundancy of the representation of the projected electron
operators in terms of the lattice spins and dopons.  This ambiguity
is removed by projecting the gauge-dependent Hamiltonian onto the
physical subspace: upon projection all the gauge-related
Hamiltonians result in one and the same physical representation. In
this way we show that the previously used and seemingly different
spin-fermion representations of the t-J model are in fact related to
each other by this projection.

The spin-dopon constraint corresponds to a Kondo-type interaction
between the lattice spins and the dopons. The dopons play the role
of the ``conduction electrons'' while the lattice spins play the
role of the ``localized magnetic moments''. This approach allows for
a mapping of the original t-J model onto a Kondo-Heisenberg lattice
model which indicates a strong relationship between the physics of
high-T$_c$ materials and that of some of the quasi two-dimensional
heavy fermion systems (see sections III-IV). We also show that this
constraint, in contrast to the standard NDO condition, can be
treated in a controlled way within mean field approximation.

Using our approach, the pseudogap phase can be simply interpreted as
a state in which the lattice spins are paired, and the low density
dopons are the quasiparticles solely responsible to generate the
small Fermi surface. If the dopon concentration increases so does
the lattice spin-dopon coupling. As a result, in the normal state
above T$_c$ there is a critical dopon concentration at which this
coupling starts to dominate over the corresponding lattice spins
self-interaction finally breaking the remaining spin pairs. When
this takes place, both spins and dopons integrate the large Fermi
surface and the physical system crosses over to a metallic phase
with no pseudogap behavior.

\section{Doped carrier representation}

The t-J model hamiltonian in terms of the Gutzwiller projected
lattice electron operators takes the form
 \be
H_{tJ}=-\sum_{ij\sigma} t_{ij} \tilde{c}_{i\sigma}^{\dagger}
\tilde{c}_{j\sigma}+ J\sum_{ij} (\vec Q_i \cdot \vec Q_j -
\frac{1}{4}\tilde{n}_i\tilde{n}_j),
\end{equation}
where $\tilde{c}_{i\sigma}=c_{i\sigma}(1-n_{i,-\sigma})$ is the
projected electron operator (to avoid the on-site double
occuppancy), $\vec
Q_i=\sum_{\sigma,\sigma}\tilde{c}_{i\sigma}^{\dagger}\vec\tau_{\sigma\sigma}\tilde{c}_{i\sigma},
\,\vec\tau^2=3/4, $ is the electron spin operator and $\tilde
n_i=n_{i\uparrow}+n_{i\downarrow}-2n_{i\uparrow}n_{i\downarrow}$.

Following Ribeiro and Wen, we consider an enlarged Hilbert space for
each site $i$. This enlarged space is characterized by the state
vectors $|\sigma a\rangle$ with $\sigma=\uparrow,\downarrow$
labelling the spin projection and $a=0,\uparrow,\downarrow$
labelling the dopon state (double occupancy is not allowed). In this
way the enlarged Hilbert space becomes \be
\mathcal{H}^{enl}=\{|\uparrow 0\rangle_i,|\downarrow
0\rangle_i,|\uparrow \downarrow\rangle_i, |\downarrow
\uparrow\rangle_i,|\uparrow \uparrow\rangle_i,|\downarrow
\downarrow\rangle_i\}
\end{equation}
while in the original Hilbert space we can either have one electron
with spin $\sigma$ or a vacancy:

\be \mathcal{H} =\{|\uparrow \rangle_i,|\downarrow
\rangle_i,|0\rangle_i\}
\end{equation}

The following mapping between the two spaces is then defined:

\begin{subequations}
\be |\uparrow \rangle_i \leftrightarrow |\uparrow 0\rangle_i
\end{equation}
\be |\downarrow \rangle_i \leftrightarrow |\downarrow 0\rangle_i
\end{equation}
\be |0 \rangle_i \leftrightarrow \frac{|\uparrow \downarrow\rangle_i
- |\downarrow \uparrow\rangle_i}{\sqrt{2}}
\end{equation}
\end{subequations}

The remaining states in the enlarged Hilbert space,
 $\frac{|\uparrow \downarrow\rangle_i + |\downarrow \uparrow\rangle_i}{\sqrt{2}}$,
$|\uparrow \uparrow\rangle_i$, $|\downarrow \downarrow\rangle_i$ are
unphysical and should be excluded in practical calculations. In this
mapping, a vacancy in the electronic system corresponds to a singlet
pair of a lattice spin and a dopon whereas the presence of an
electron is related to the absence of a dopon.

Let us call $\vec S_i$ the operator associated with the lattice spin
on site $i$ and $d_{i\sigma}^{\dagger}$ the creation operator for
dopons. Then under this mapping it is not difficult to find
relations between the original projected electronic operators and
the new operators, such as: \be \label{eq:c} \tilde{c}_{i\uparrow}=
\frac{1}{\sqrt{2}} \left[ S_i^+ S_i^-
\tilde{d}_{i\downarrow}^{\dagger} -S_i^-
\tilde{d}_{i\uparrow}^{\dagger}   \right]
\end{equation}
where
$\tilde{d}_{i\sigma}=d_{i\sigma}(1-d_{i,-\sigma}^{\dagger}d_{i,-\sigma})$
is a projected dopon operator. Although $\vec S_i$ and
$\tilde{d}_{i\sigma}$ act in the whole enlarged Hilbert space,
specific combinations like the one given in Eq. (\ref{eq:c})
nullifies the unphysical states. It is then possible to write the
original t-J Hamiltonian in terms of the new operators in such way
that it vanishes when acting on the unphysical states~
\cite{ribeiro06}. The reason for this construction is now obvious:
close to half-filling there is a small amount of dopons in the
system and as a result the local constraint of no double dopon
occupancy can be safely dropped.

However, as soon as some mean-field approximations are performed,
the unphysical states reappear in the theory \emph{regardless of
whether projected or unprojected dopon operators are used}. A
constraint should therefore be imposed in order to eliminate the
unphysical states. This constraint was proposed in ~\cite{ferraz07}
and it reads \be \label{eq:constr} \hat{\Upsilon}_i=\vec{S_i} \cdot
\vec{M_i}+\frac{3}{4}(\tilde{d}_{i\uparrow}^{\dagger}\tilde{d}_{i\uparrow}+
\tilde{d}_{i\downarrow}^{\dagger}\tilde{d}_{i\downarrow})=0
\end{equation}
where $\vec
M_i=\sum_{\sigma,\sigma}\tilde{d}_{i\sigma}^{\dagger}\vec\tau_{\sigma\sigma}\tilde{d}_{i\sigma}
$  is the spin associated with the dopon. Acting on the physical
states, $\hat{\Upsilon}_i$ gives zero while $\hat{\Upsilon}_i
|unphys\rangle_i =|unphys\rangle_i$.

Since for $1/2$ spin $\vec S^2_i=3/4$ and $\vec M_i^2=(3/4)\tilde
n^d_i$, this constraint can also be recast into the form
$$\vec J_i^2-3/4(1-\tilde n^d_i)=0.$$
Here $ \vec J_i=\vec M_i+\vec S_i$ is the total spin on each lattice
site. In this form the constraint tells us that the on-site total
spin can be either $j=0$ (vacancy) or $j=1/2$ (lattice spin).

There is a yet another representation of this constraint, namely
\be\hat\Upsilon_i=-\frac{1}{2}D^{\dagger}_i D_i +\tilde n^d_i,
\label{costr3}\end{equation} where
$D^{\dagger}_i=f^{\dagger}_{\uparrow i}\tilde
{d}^{\dagger}_{\downarrow i}-f^{\dagger}_{\downarrow i}\tilde
{d}^{\dagger}_{\uparrow i}$ creates an on-site vacancy. Here we
represent the spin degrees of freedom in terms of chargeless
fermions (spinons), $\vec S_i=f^{\dagger}_i\vec\tau f_i, \quad
f_i=(f_{\uparrow i}, f_{\downarrow i})^t, \quad f^{\dagger}_if_i=1.$
In this form the constraint $\sum_i\hat\Upsilon=0$ tells us that the
total number of vacancies must be equal to the total number of
dopons \cite{china}.

Right at half-filling $(\tilde n^d_i=0)$, the spin-dopon constraint
becomes a trivial identity, $0=0$. It can thus be safely treated
close to half-filling at a mean-field level. In contrast, in the
overdoped regime $(\tilde n^d_i\approx 1)$ it reduces to the
equation, $\vec{S_i} \cdot \vec{M_i}\approx -3/4.$ Exactly the
opposite situation is realized for the lattice electron NDO
constraint: it cannot be relaxed close to half-filling but it can be
totally ignored for a low enough density of the lattice electrons.
It is precisely because of this that the new dopon-spin t-J model
representation becomes indeed quite appealing.

It is convenient to define the operator
$\hat{\mathcal{P}}_i=1-\hat{\Upsilon}_i$. Since
$\hat{\Upsilon}^2_i=\hat{\Upsilon}_i$ it is clear that
$\hat{\mathcal{P}}_i$ is a projection operator which eliminates
unphysical states on the site $i$ \cite{p}. We show now that by
means of this projection operator we can indeed establish
alternative ways of expressing the electron operators in terms of
the lattice spins and dopons. Let us consider for example the action
of $\tilde{d}_{i\downarrow}$ on the physical states -
 it is simply $\tilde{d}_{i\downarrow} |\uparrow 0\rangle_i = 0$,
 $\tilde{d}_{i\downarrow}|\downarrow 0\rangle_i = 0$ and
$\tilde{d}_{i\downarrow} \frac{|\uparrow \downarrow\rangle_i -
|\downarrow \uparrow\rangle_i}{\sqrt{2}}= \frac{|\uparrow
0\rangle_i}{\sqrt{2}}$. Clearly, in the physical subspace this is
equivalent to the action of
$\tilde{c}_{i\uparrow}^{\dagger}/\sqrt{2}$. Thus we can readily
write: \be  \tilde{c}_{i\uparrow}^{\dagger}= \sqrt{2}
\hat{\mathcal{P}}_i  \tilde{d}_{i\downarrow}
\hat{\mathcal{P}}_i\label{eq:7}
\end{equation}

Alternatively we can use the projection operator
$\hat{\mathcal{P}}_i$ to perform  explicit calculations, resulting
in
$$ \sqrt{2} \hat{\mathcal{P}}_i  \tilde{d}_{i\downarrow}
\hat{\mathcal{P}}_i = \frac{1}{\sqrt{2}} \left[ S_i^+ S_i^-
\tilde{d}_{i\downarrow} -S_i^+ \tilde{d}_{i\uparrow} \right]$$ which
combined with the adjoint of Eq. (\ref{eq:c}) results again in our
Eq. (\ref{eq:7}). In the same way one can show that
\begin{equation}
\tilde{c}_{i\downarrow}^{\dagger} = -\sqrt{2} \hat{\mathcal{P}}_i
\tilde{d}_{i\uparrow} \hat{\mathcal{P}}_i.\label{9}\end{equation}

Consider now the action of $S_{i}^{+} \tilde{d}_{i\uparrow}$ on the
physical states, namely, $S_{i}^{+}\tilde{d}_{i\uparrow} |\uparrow
0\rangle_i = 0$, $S_{i}^{+}\tilde{d}_{i\uparrow} |\downarrow
0\rangle_i = 0$ and $S_{i}^{+}\tilde{d}_{i\uparrow} \frac{|\uparrow
\downarrow\rangle_i - |\downarrow \uparrow\rangle_i}{\sqrt{2}}=-
\frac{|\uparrow 0\rangle_i}{\sqrt{2}}$. Comparing with the action of
$\tilde{c}_{i\uparrow}^{\dagger}/\sqrt{2}$, or carrying out the
explicit calculations with the ${\mathcal{P}}$ projection we get yet
another operator identity: \be \label{eq:cdag2}
\tilde{c}_{i\uparrow}^{\dagger} \equiv -\sqrt{2} \hat{\mathcal{P}}_i
S_{i}^{+}\tilde{d}_{i\uparrow} \hat{\mathcal{P}}_i,
\end{equation}
whereas from the action of $S_{i}^{-} \tilde{d}_{i\downarrow}$ we
immediately find \be \label{eq:cdag3}
\tilde{c}_{i\downarrow}^{\dagger}= \sqrt{2} \hat{\mathcal{P}}_i
S_{i}^{-}\tilde{d}_{i\downarrow} \hat{\mathcal{P}}_i
\end{equation}
Note that a few attempts have also been made at decoupling the
physical electron as the spinful fermion and spinon. However, they
can only be justified within some approximation scheme (see, e.g.,
ref.\cite{feng1}). In contrast, here we display the exact form of
the spin-fermion decoupling of the projected lattice electron
operator in Eqs. (10)-(11). In the next section we show that the
seemingly different representations for one and the same lattice
electron operator (e.g., given by Eqs. (9) and (11)) are in fact
identical since the ${\mathcal{P}}$ projections of gauge related
objects coincide with each other.

\section{Local gauge symmetry}

Let us now define the global projection operator
$\hat{\mathcal{P}}=\prod_i \hat{\mathcal{P}}_i$. The t-J Hamiltonian
can be generally written as $H_{tJ}=\hat{\mathcal{P}} \tilde{H}
\hat{\mathcal{P}}$, where $\tilde{H}$ acts on the enlarged Hilbert
space. From the discussion above it is clear that there are
different choices of $\tilde{H}$ that after being projected onto the
physical subspace result in the t-J model Hamiltonian. This is
related to the gauge ambiguity of  the unprojected $\tilde{H}$ with
respect to the local U(1) gauge transformations generated by the
constraint $\hat{\Upsilon}_i.$ Under this transformation,
$$\tilde d_{\uparrow i}\to \tilde d_{\uparrow i}(\theta_i)=\exp(-i\theta_i\hat{\Upsilon}_i)\tilde d_{\uparrow i}
\exp(i\theta_i\hat{\Upsilon}_i)$$
\begin{equation}
=\frac{1}{2}\left[S^{-}_i\tilde d_{\downarrow
i}+(\frac{3}{2}+S^z_{i})\tilde d_{\uparrow
i}\right](e^{i\theta_i}-1) +\tilde d_{\uparrow i},\label{1}
\end{equation}
\begin{equation}
\tilde d_{\downarrow i}(\theta_i)= \frac{1}{2}\left[S^{+}_i\tilde
d_{\uparrow i}+(\frac{3}{2}-S^z_{i})\tilde d_{\downarrow
i}\right](e^{i\theta_i}-1) + \tilde d_{\downarrow i}. \label{2}
\end{equation}
The explicit form of the gauge dependent operator $\vec S_i(\theta)$
is given by
\begin{eqnarray}
\vec S_i(\theta_i)&=&[\vec S_i\times \vec
M_i]\sin\theta_i\nonumber\\
&+&\frac{\vec S_i\tilde n_i^d-\vec M_i}{2}(\cos\theta_i-1)+\vec S_i.
\label{2a}\end{eqnarray} Here $\tilde d_{\sigma i}\equiv \tilde
d_{\sigma i}(\theta=0), \, \vec S_i\equiv \vec S_i(\theta=0).$ The
U(1) gauge symmetry is realized on the spin-dopon multiplet in a
nontrivial way. Note also that the total on-site electron spin
operator $\vec J_i=\vec S_i+\vec M_i$ as well as the dopon number
operator are gauge invariant quantities, $\vec J_i(\theta)=\vec
J_i(\theta=0), \, \tilde n^d_i(\theta)=\tilde n^d_i(\theta=0).$

The $\hat{\mathcal{P}}$-projection of the different gauge equivalent
operators results in the same gauge invariant representation. For
instance, for any polynomial on-site operator $\hat f_i=\hat
f_i(\tilde d_{\sigma i}, S_i)$ we get
$$\hat{\mathcal{P}}_i\hat f_i(\theta)
\hat{\mathcal{P}}_i=
\hat{\mathcal{P}}_i\exp(-i\theta_i\hat{\Upsilon}_i)\hat
f_i\exp(i\theta_i\hat{\Upsilon}_i) \hat{\mathcal{P}}_i$$
$$=\hat{\mathcal{P}}_i\!\left[1+\hat{\mathcal{P}}_i(e^{i\theta_i}-1)\left]\hat
f_i\right[1+\hat{\mathcal{P}}_i(e^{-i\theta_i}-1)\right]\!\hat{\mathcal{P}}_i=\hat{\mathcal{P}}_i\hat
f_i\hat{\mathcal{P}}_i.$$ In this way the representation (11)
follows from the equations
$$\hat{\mathcal{P}}_i\tilde d_{\uparrow
i}\hat{\mathcal{P}}_i= \hat{\mathcal{P}}_i\tilde d_{\uparrow
i}(\theta=\pi)\hat{\mathcal{P}}_i=-\hat{\mathcal{P}}_i(S^{-}_i\tilde
d_{\downarrow i})\hat{\mathcal{P}}_i$$
$$=\frac{1}{2}\left[S^{-}_iS^{+}_i\tilde d_{i\uparrow}-S^{-}_i\tilde
d_{i\downarrow}\right].$$ Notice that in general
$$\hat{\mathcal{P}} \sum_{ij\sigma} t_{ij}
\tilde{d}_{i\sigma}^{\dagger}(\theta_i)
\tilde{d}_{j\sigma}(\theta_j)\hat{\mathcal{P}} = \hat{\mathcal{P}}
\sum_{ij\sigma} t_{ij} \tilde{d}_{i\sigma}^{\dagger}
\tilde{d}_{j\sigma}\hat{\mathcal{P}}.$$ It can also be checked by an
explicit computation that the projected electron operators, e.g.,
the ones given by Eqs. (\ref{eq:7}-\ref{eq:cdag3}) are all gauge
invariant objects. In analogy with the gauge theories from quantum
field theory, we can say that essentially the choice of a given
representation for $\tilde{H}$ corresponds to the fixing of a
particular gauge.

The existence of the local gauge symmetry reflects a degree of
redundancy in the parametrization of the Gutzwiller projected
lattice electrons in terms of the lattice spins and dopons, as
displayed in our Eq.~(\ref{eq:c}). In principle, one can formulate a
mean-field theory in this representation that respects that local
U(1) gauge symmetry. In this way one arrives at a local gauge theory
that describes quantum fluctuations around the mean-field solution.

It should be kept in mind that the lattice spins and dopons are, in
general, not gauge invariant and couple to the gauge field. Because
of this the dopons and lattice spins, away from half-filling, do not
represent real excitations and they are introduced as an
intermediate step to calculate the physical (gauge-invariant)
quantities such as given, e.g., by Eqs.~(\ref{eq:c}), (8)-(11). Note
however that, right at half-filling, $\vec S_i\to \vec S_i$. In
other words the lattice spins represent, in this limit, real
excitations. It is therefore natural to assume that close to
half-filling the lattice spins and dopons can be viewed as
well-defined excitations weakly coupled to the gauge field. This indicates 
that the mean-field spin-dopon theory
\cite{comment} is presumably stable close to half-filling with
respect to quantum gauge fluctuations. However, an explicit
estimation of the strength of the gauge interaction can be made only
after the full gauge theory is derived. The explicit form of that
theory is not still available because of a rather complicated form
of the U(1) group action on the spin-dopon multipletes (see Eqs.
(\ref{1}-\ref{2a})).

To make this point more clear, let us contrast the properties of the
gauge symmetry generated by $\hat{\Upsilon}_i$ with those of the
U(1) local gauge symmetry generated by the standard NDO constraint
in the frequently used slave-boson representation for the lattice
electron operators. With this formalism the projected electron
operator is written as
\begin{equation}
\tilde c_{i\sigma}^{\dagger}=f^{\dagger}_{i\sigma}b_i,
\label{sb}\end{equation} with the NDO condition
$$\hat Q_i:=\sum_{\sigma}f^{\dagger}_{i\sigma}f_{i\sigma}+b^{\dagger}_ib_i=1,$$
where $f_{i\sigma}$ is a fermion (spinon) operator and $b_i$ is a
slave-boson (holon) operator. Conservation of the gauge charge $\hat
Q_i$ can be derived by the Noether theorem starting from the local
U(1) gauge transformation, \cite{lnw}
\begin{equation}
f_{i\sigma}\to e^{i\theta_i}f_{i\sigma}, \quad b_i\to
e^{i\theta_i}b_i, \label{sbtr}\end{equation} which leaves the
physical electron operator (\ref{sb}) intact. This U(1) local gauge
symmetry takes care of the redundancy of the parametrization
(\ref{sb}). In contrast to the spin-dopon charge $\hat\Upsilon_i$,
the operator $\hat Q_i$ does not vanish at half-filling, $\hat
Q_i\to \hat Q_i^{hf}=
\sum_{\sigma}f^{\dagger}_{i\sigma}f_{i\sigma}.$   This indicates that
the spinons are strongly coupled to the bare gauge field close to half filling. However,
the auxiliary gauge field can in general acquire nontrivial dynamics at low energies, effectively moving 
the model to a weak coupling regime. Therefore, to judge whether confinement or deconfinement 
of slave particles really occurs in the physical low-energy excitations, one must explicitly 
investigate the gauge dynamics in the low-energy regime \cite{ichinose}.

Note also that in contrast with the NDO constraint for the lattice
electrons, the set of local constraints $\hat{\Upsilon}_i=0$ (one
for each lattice site) can be replaced by the global condition
$\hat{\Upsilon}=\sum_i \hat{\Upsilon}_i=0$. The reason for this
simplification is that the unphysical states appear as the
degenerate eigenvectors of $\hat{\Upsilon}_i$ with an eigenvalue, 1.
Therefore, acting on an unphysical state, $\hat{\Upsilon}$ simply
produces the same state multiplied by a positive number. Acting on a
physical state, $\hat{\Upsilon}$ always gives zero. Note, however,
that $\hat{\Upsilon}_i$ involves a quartic power of interacting
dopons and spins. The standard NDO constraint appears as a quadratic
form of the electron operators. We will comment on this point
further at the end of the paper.

Summing up all the above, we can write down the exact form of the
t-J Hamiltonian in the spin-dopon representation as

$$H_{tJ}=\hat{\mathcal{P}}  \sum_{ij\sigma} 2t_{ij}
\tilde{d}_{i\sigma}^{\dagger} \tilde{d}_{j\sigma}\hat{\mathcal{P}}$$
\begin{equation}
+ J\sum_{ij}\hat{\mathcal{P}}((\vec S_i + \vec M_i)(\vec S_j + \vec
M_j)-\frac{1}{4}(1-\tilde{n}^d_i)(1-\tilde{n}^d_j))\hat{\mathcal{P}}
\label{tj1}\end{equation} Within the path-integral approach this
representation has been used in~\cite{ferraz06} to obtain the
mean-field $T_c$ phase diagram for the cobaltates. The exact
spin-dopon path-integral representation of the t-J partition
function given in~\cite{ferraz06} is written down in terms of the
classical fermion amplitudes $\psi_{i\sigma}$ which are related to
the dopon amplitudes in the following way,
$\psi_i=\frac{1}{\sqrt{2}}(\frac{1}{2}-2\vec S^{cl}_i\vec\tau)\tilde
d_i^{cl}=\sqrt{2}(\hat{\mathcal{P}_i}
\tilde{d}_i\hat{\mathcal{P}_i})^{cl}$. This relation holds provided
$\Upsilon^{cl}_i=0$. Note that in this case
$(n^{\psi}_i)^{cl}=(\tilde n_i^{d})^{cl}.$ Within that path-integral
approach the constraint to exclude the unphysical states reasserts
itself in the form of the SU(2) invariant site product of the
delta-functions that singles out the physical subspace.

Since $\vec{Q}_i=\hat{\mathcal{P}}_i (\vec S_i + \vec
M_i)\hat{\mathcal{P}}_i = \vec S_i (1-\tilde
d_{i\sigma}^{\dagger}\tilde d_{i\sigma})$, and $\hat{\mathcal{P}}_i
(1-\tilde d_{i\sigma}^{\dagger}\tilde
d_{i\sigma})\hat{\mathcal{P}}_i=(1-\tilde
d_{i\sigma}^{\dagger}\tilde d_{i\sigma})$, we can rewrite
Eq.(\ref{tj1}) in the form
$$H_{tJ}=\hat{\mathcal{P}}  \sum_{ij\sigma} 2t_{ij}
\tilde{d}_{i\sigma}^{\dagger} \tilde{d}_{j\sigma}\hat{\mathcal{P}}$$
\begin{equation}
+ J\sum_{ij} (\vec S_i \cdot \vec S_j -
\frac{1}{4})(1-\tilde{n}^d_i) (1-\tilde{n}^d_j). \label{tj2}
\end{equation}
Note the important factor of $2$ in front of the $t$-term in these
formulas. It originates  from the fact that the vacancies are
represented in this theory by the spin-dopon singlets given by
Eq.(4c).

The $\hat{\mathcal{P}}$-projected dopons describe the physical doped
carriers. Calculating explicitly the $\hat{\mathcal{P}}$-projected
dopon operators  we get
$$H_{tJ}=  \sum_{ij\sigma}\frac{t_{ij}}{2}\,
\tilde{d}_{i}^{\dagger}(\frac{1}{2}-2\vec
S_i\vec\tau)\,(\frac{1}{2}-2\vec S_j\vec\tau)\, \tilde{d}_{j}$$
\begin{equation}
+ J\sum_{ij} (\vec S_i \cdot \vec S_j -
\frac{1}{4})(1-\tilde{n}^d_i) (1-\tilde{n}^d_j). \label{tj3}
\end{equation}
The representation~(\ref{tj3}) has been used by Ribeiro and Wen
within the mean-field approximation. However, they oversimplified
the magnetic term in the following way,
\begin{eqnarray}
&&J\sum_{ij} (\vec S_i \cdot \vec S_j -
\frac{1}{4})(1-\tilde{n}^d_i) (1-\tilde{n}^d_j)\to \nonumber\\
&& \tilde J\sum_{ij} (\vec S_i \cdot \vec S_j - \frac{1}{4}),
\label{tj4}
\end{eqnarray}
where $\tilde{J}=J(1-x)^2$, $x$ is a density of dopons. It is clear
that the representation (\ref{tj4}) totally ignores the dynamically
induced doping changes in the underlying spin correlations. The
authors instead introduce phenomenological $x$-dependent hopping
parameters to take into account the feedback of the dopons on the
spin dymanics. To take into account the actual $J$ dependent
spin-dopon interaction, one should use the constraint
$\hat{\Upsilon}_i=0.$ It is satisfied provided $\vec M_i+\tilde
n_i\vec S_i=0.$ This yields (up to unessential constant factors),
$$H_{tJ}=  \sum_{ij\sigma}\frac{t_{ij}}{2}\,
\tilde{d}_{i}^{\dagger}(\frac{1}{2}-2\vec
S_i\vec\tau)\,(\frac{1}{2}-2\vec S_j\vec\tau)\, \tilde{d}_{j}$$
\begin{eqnarray}
&+& J\sum_{ij}\left[ (\vec S_i \cdot \vec S_j -1/4) +(\vec S_i\vec
M_j+ \vec S_j\vec M_i)\right.\nonumber\\
&&\left. +(\vec M_i\vec M_j-\tilde n_i\tilde
n_j/4)\right]\label{tj5}
\end{eqnarray}
The magnetic term in this representation explicitly accounts for the
spin-dopon interaction as produced by the magnetic moment-moment
interactions. In general, it is the representation (\ref{tj5}) that
should be used as a starting point to apply a mean-field
approximation. In this way a complete dynamical mean-field phase
diagram to describe hole/electron doped cuprates emerges in contrast
to a semi-phenomenological one previously derived
\cite{ribeiro05,ribeiro06}.

Close to half-filling an alternative way of dealing with the
constraint can be proposed. First, we can drop the projection
operator and add the constraint with an appropriate Lagrange
multiplier. That is, we now write:
$$H_{tJ}= \sum_{ij\sigma} 2t_{ij}
\tilde{d}_{i\sigma}^{\dagger} \tilde{d}_{j\sigma}$$
\begin{equation}
+ J\sum_{ij} (\vec S_i \cdot \vec S_j -
\frac{1}{4})(1-\tilde{n}^d_i) (1-\tilde{n}^d_j)+ \lambda
\hat{\Upsilon}, \label{tj6}
\end{equation}
where $\lambda$ is to be send to $+\infty$ at the end of
calculations. Second, since the dynamics is now restricted  to the
physical subspace,  we can close to half-filling make the change
$J\to \tilde J$ to get \be \label{eq:H1} H_{tJ} = \sum_{ij\sigma}
2t_{ij} {d}_{i\sigma}^{\dagger} {d}_{j\sigma}+ \tilde{J}\sum_{ij}
(\vec S_i \cdot \vec S_j - \frac{1}{4}) + \lambda \hat{\Upsilon}.
\end{equation}
By writing the Hamiltonian in this way, we see the decisive role
played by the constraint which incorporates the interaction between
dopons and lattice spins, while the non-constrained Hamiltonian
simply corresponds to non-interacting dopons and lattice spins.

The large-$\lambda$ limit eliminates unphysical states with the
total spin $j=1$ and at the same time dynamically generates a
vacancy on a lattice site in the form of a spin-dopon
singlet~\cite{ferraz07}. An analogy of this result can be drawn with
the dynamical formation of the Zhang-Rice singlet produced by the
hybridization effects that strongly bind a hole and a Cu$^{2+}$ ion
together to form a local singlet state associated with such a
vacancy \cite{zr}. The important point here being again the fact
that such an empty site (vacancy) can be physically interpreted as a
spin singlet.

We now introduce the chemical potential for dopons and use the
explicit representation of the constraint, Eq. (\ref{eq:constr}), to
get the Hamiltonian: \be \label{eq:Hf} H_{tJ} = \sum_{ij\sigma}
t^{eff}_{ij}{d}_{i\sigma}^{\dagger} {d}_{j\sigma}+
\tilde{J}\sum_{ij} (\vec S_i \cdot \vec S_j - \frac{1}{4}) +\lambda
\sum_{i}\vec{S_i} \cdot \vec{M_i}
\end{equation}
where $t^{eff}_{ij}=2t_{ij}+(\frac{3}{4}\lambda-\mu) \delta_{ij}$.
The parameter $\lambda$ must be sent to $+\infty$ at the end of
calculations \cite{lambda}.

Finally, we can safely treat the constraint close to half-filling at
the mean-field level. In this case, $\lambda$ is determined
self-consistently from the ground-state average $$<\sum_i(\vec{S_i}
\cdot \vec{M_i}+\frac{3}{4}\tilde n_i)> =0$$ and becomes doping
dependent. The mean-field Hamiltonian obtained in this way is that
of a Kondo-Heisenberg lattice problem, where the lattice spins play
the role of localized magnetic moments while the dopons take the
role of conduction electrons.

In order to take into account a possible hybridization between the
localized spins and the dopons \cite{paul07} one should use the
following form of the constraint
$$<-\frac{1}{2}\sum_iD^{\dagger}_i D_i +\sum_i\tilde n^d_i>=0.$$
The relevant order parameter takes the form $\sigma:=<D>$, where we
have linearized the on-site operator product in the averaged
constraint in the following way, $D^{\dagger}_i D_i\approx
D^{\dagger}_i <D> + <D^{\dagger}>D_i -<D^{\dagger}><D>.$ This
procedure preserves the SU(2) symmetry, which must be present in the
underdoped phase. Since $\sigma \propto \sqrt{x},$ the error
produced by this simplification is at most of order ${\cal O}(x),
x\to 0,$ which does not affect the results quantitatively in that
regime. The breakdown of the Kondo regime implies then $\sigma=0.$

\section{Physical Implications}

In the previous section we showed that the t-J model can be mapped
onto a Kondo-Heisenberg model for dopons and lattice spins. In this
section we explore some immediate physical implications of that
mapping, leaving a more detailed analysis to a future work.

The Kondo-Heisenberg lattice model has attracted much attention in
the context of heavy-fermion systems
\cite{tsunetsugu97,senthil04,fulde06,paul07}. It is believed that a
variety of physical phenomena could be  accounted for by that model,
such as non-Fermi liquid behavior, different types of magnetic,
charge ordering and perhaps unconventional superconductivity
\cite{coleman}.

In fact, recent experiments have revealed striking similarities
between quasi two-dimensional heavy fermion systems (the
Ce\emph{M}In$_5$ family) and the high-$T_c$ cuprates
\cite{sidorov02,bel04,nakajima07}. The mapping discussed in the last
section suggests that these similarities could be accounted for by
the fact that both quasi-$2d$ heavy fermions and high-T$_c$ cuprates
capture universal features of strongly correlated electron systems
in the presence of strong AF correlations.

Our mapping reinforces earlier suggestions of a common magnetically
mediated mechanism for superconductivity in heavy fermion compounds
and in the cuprates \cite{mathur98,moriya03}. In this way, the
superconductivity in the cuprates can be directly associated with
the pairing of dopons induced by the Kondo like interaction with the
lattice spins. However, if this is indeed the case, a crucial
question arises naturally: why are the critical superconducting
transition temperatures observed in heavy fermions (low T$_c$) and
cuprates (high T$_c$) so different from each other? Our explanation
for that is signalled by the different charge carrier mass
renormalizations and the typical coupling constant magnitudes in
those two referred systems. Suppose the critical superconducting
temperature is given generically by $T_c=\Delta
\exp{(-cm^*/\lambda)}$ where $c$ is a constant, $m^*$ is the charge
carrier effective mass, $\lambda$ is the lattice spin-dopon coupling
and $\Delta$ is some typical energy scale. In heavy fermion
compounds $\lambda= J_K$ is small, and $m^*$ is two or three orders
of magnitude bigger than the bare electron mass. This leads to a
very small T$_c$. In contrast, from infrared Hall measurements on
underdoped LSCO and YBCO, $m^*$ is of the same order of magnitude as
bare electron mass and $\lambda$ is large. This is due to the fact
that in the large $\lambda$ limit, for optimally doped cuprates, we
arrive at a Kondo like regime with $T_c\propto T_K\propto D$, the
dopon bandwidth.

Let us now connect our result more directly to the recent
experimental results of Doiron-Leyraud et al \cite{3}. Right at
half-filling and below Neel temperature T$_N\propto J/4$, the
antiferromagnetic ordering is accounted by the Heisenberg
interaction term in Eq. (\ref{eq:Hf}). Above T$_N$, thermal
fluctuations destroy the long-range order. However, since the spin
exchange energy $J$ is in fact extraordinarily large the system
still shows strong short-range AF correlations well above T$_N$.
This phase is accounted for by the spin liquid state of the
spin-spin singlets. As the dopon concentration increases, the
long-range AF order is melted by the quantum mechanical jiggling of
the local spin moments induced by the small (in this regime)
dopon-spin interaction, and it eventually disappears altogether.
Although the RKKY spin-spin interaction induced by dopons produces
by itself the long-range AF ordering of the lattice spins, its
strength is $\propto \lambda ^2$ which is much less than the
spin-spin exchange energy, $J$. As a result, at some finite dopon
concentration, the AF long-range order gives way to short-range AFM
spin-spin correlations and the lattice spins become paired. As the
doping increases, the individual lattice spins become less
correlated with each other due to the competition between AFM
fluctuations and the Kondo screening.

Suppose we are now in the pseudogap regime. The lattice spins form
singlet pairs interacting with the dopons by means of a still weak
$\lambda$ coupling. The low density dopons are the only fermionic
carriers present in the system which can be associated with the
small hole pocket Fermi surface (FS) of the pseudogap state. The
small volume of such a FS is accounted by the low density dopons
present in the system. As the density of dopons increases, the
dopon-spin coupling also increases and the dopon-spin singlets
evolve continuously out of the pseudogap state into a more Kondo
like regime \cite{wen HF}.

Let us now estimate the critical density associated with such a
crossover. The necessary energy to break the lattice spin pairs is
roughly $J$. Since the individual spins in each pair and dopons
become closely coupled to each other by means of the increase of
$\lambda$ , the dopon kinetic energy $2tx$, where $x$ is dopon
concentration, soon becomes of the order of $J$. Consequently, when
$2tx>J$, the spin gap is destroyed and the resulting FS is now
enlarged by the presence of the highly correlated charged spins
which together with the dopons are now counted as charge carriers.
Taking simply $J/t\approx 1/3$, we arrive immediately at the lower
bound estimate for the critical density $x_c\approx 0.17$, which is
in very good agreement with the experimental value for the small
versus large FS crossover which, for the hole doped cuprates takes
place in the doping range of $0.15-0.25$, and it is associated with
the complete disappearance of the pseudogap state.

It is also worthwhile at this stage to compare our scheme with the
earlier mean-field slave-boson formulation.  In that representation,
the electron operator is decoupled into  a spinon (fermion) and a
holon (boson). Clearly when the spinons are paired into singlet
states and the charged bosons are not yet Bose-Einstein condensed
the resulting spin gapped state has no FS to be associated with.
This is in direct disagreement with the recent experimental findings
which demonstrate the metallic character of the pseudogap phase.

To complete the overall discussion of the physical implications of
our dopon-lattice spin system, we need to clarify the onset and
disappearance of the superconducting state in both underdoped and
overdoped regimes. In the underdoped limit, as emphasized earlier,
the lattice spins form singlet pairs interacting weakly with the
dopons through $\lambda$. This interaction $\lambda$ naturally leads
dopons to condense into BCS like pairs at temperature $T\leq T_c$.
Notice that since the coupling $\lambda$ is weak at low doping, the
superconducting gap resulting from BCS condensation of dopons is
strongly doping dependent and it is not directly related to the spin
gap. As a result the superconducting gap and the spin gap
(pseudogap) are, in practice, two independent energy scales at very
low doping. This is in agreement with recent angle-resolved
photoemission (ARPES) and Raman experiments which distinguish the
roles of the nodal and anti-nodal gaps in the low doping
superconducting phase. In contrast, in the optimal doping region for
larger dopon concentration, the NDO constraint must be treated with
care and as a result $\lambda$ grows accordingly. Such a growth of
$\lambda$ strongly ties the spin and dopon to each other. Therefore,
at sufficiently large dopings, the spin and dopon gaps should become
indistinguishable from each other.  The superconducting phase is
well described in this regime by a single energy scale. As a result
at a sufficiently large $\lambda$, both gaps are destroyed
simultaneously and we end up with a low-energy Kondo like metal
\cite{cobalt}. Needless to say, the projection NDO constraint is a
crucial ingredient in all our arguments and it allows us to give a
simple explanation of important recent results.

\section{Conclusion}

In the present paper we discuss the physical meaning and some
implications of the theory of the projection constraint in the doped
carrier representation of the t-J model. The basic conclusions that
can be drawn from our consideration are as follows. Firstly, the
complete theory that incorporates the constraint sounds quite
appealing, since it allows for a controlled mean-field treatment of
the t-J model in the most interesting region close to half-filling.
This happens at the expense of having more complicated interaction
terms in the spin-dopon t-J Hamiltonian. It would be of utmost
importance to derive the complete mean-field theory in this
representation which is now under consideration. However, as a first
step in that direction we just discuss some immediate qualitative
consequences of the improved spin-dopon approach.

Secondly, the constraint enforced by the Lagrange multiplier term
allows for an explicit mapping of the t-J model onto the
Kondo-Heisenberg lattice model in the underdoped region . This
indicates that the physics behind these two model are indeed related
to each other. This mapping is very appealing in view of recent
experiments that suggest striking similarities between quasi two
dimensional heavy fermion systems and high-$T_c$ cuprates. Some
physical implications are briefly discussed, pointing to the unified
physics of heavy fermion and high-$T_c$ materials. Namely, it is
possible that the very same physical mechanism is responsible for
the formation of Cooper pairs in those systems, with different
critical temperatures related to different mass renormalization of
the charge carriers and to different magnitudes of the existing
coupling constants.

We also discuss the small-large evolution of the FS with doping. We
associate this crossover to the closure of the spin gap and the
destruction of the pseudogap state. We estimate the lower bound
density for such a crossover. We make direct contact with the recent
FS experiment of Doiron-Leyraud et al. When the spin gap is present,
the pseudogap state has a small FS with the dopons being the only
available charge carriers. With the increase of the dopon
concentration and with the consequent increase of the coupling
$\lambda$, the dopons and the individual lattice spins become
strongly correlated to each other. In this way when the spin gap is
destroyed, above the superconducting temperature T$_c$, both dopons
and lattice spins generate the associated FS.

Making explicit use of the projection constraint we also discuss the
onset and disappearance of superconductivity in both underdoped and
overdoped regimes. Further work is however needed to explore in a
more quantitative basis our new mean-field doped carrier formulation
of the t-J model. This is already in progress and the results will
be presented elsewhere.

\section{Acknowledgments}

We thank S. Malik and M. Sigrist for stimulating discussions. This
work was partially supported by the Brasilian Ministry of Science
and Technology and by CNPq.

\section{Appendix: Analogy with the Gutzwiller projection}

The global projection operator described above is similar to that
used in the $U=\infty$ Hubbard model. Let us consider the
Hamiltonian $H_{Hub} = \sum_{ij\sigma} t_{ij} c_{i\sigma}^{\dagger}
c_{j\sigma} + U\sum_i n_{i\uparrow}n_{i\downarrow}$. In the case $U
\to \infty$,  the system is subject to the constraint
$n_{i\uparrow}+n_{i\downarrow} \le 1$. This constraint is equivalent
to $\hat\Upsilon^{G}_i=n_{i\uparrow}n_{i\downarrow}=0$. In this way
when $\hat\Upsilon^{G}_i$ acts on the unphysical state (doubly
occupied) we have $\hat\Upsilon^{G}_i
|unphys\rangle_i=|unphys\rangle_i$. Therefore, $\hat
P_i^{G}=1-n_{i\uparrow}n_{i\downarrow}$ is a projection operator
that eliminates the unphysical state at site $i$. The gauge
transformation generated by this constraint,
$$c_{\downarrow}\to c_{\downarrow}e^{i\epsilon n_{\uparrow}}, \,
c_{\uparrow}\to c_{\uparrow}e^{i\epsilon n_{\downarrow}},$$ leaves
the projected electron operators $\tilde
c_{\sigma}=c_{\sigma}(1-n_{-\sigma})$ intact. The global projection
operator is the well known Gutzwiller projector $\hat P^{G}=\hat
\Pi_i \hat P_i^{G}$. We can then implement the constraint writing
$H_{Hub} =\hat P^{G} \sum_{ij\sigma} t_{ij} c_{i\sigma}^{\dagger}
c_{j\sigma}\hat P^{G}$, which is equivalent to $H_{Hub}
=\sum_{ij\sigma} t_{ij} \tilde{c}_{i\sigma}^{\dagger}
\tilde{c}_{j\sigma}$. From this point of view, the parameter $U$ of
the Hubbard model becomes the Lagrange multiplier which is necessary
to enforce the NDO constraint.


\begin{thebibliography}{10}

\bibitem{1} K.M. Shen et al, Science {\bf 307}, 901 (2005).

\bibitem{2} T. Timusk and B. Staff, Rep. Prog. Phys. {\bf 62}, 61
(1999).

\bibitem{3} N. Doiron-Leyraud et al, Nature {\bf 477}, 565 (2007).

\bibitem{4} U. Chatterjee et al, Phys. Rev. Lett. {\bf 96}, 107006
(2006).

\bibitem{5} A. Damascelli et al, Rev. Mod. Phys. {\bf 75}, 473
(2003).

\bibitem{6} J.C. Campuzano et al, in "The Physics of
Superconductors" vol. 2, ed. by K.H. Bennemann and J.B. Ketterson,
Springer (2004).

\bibitem{7} See e.g., D. Zanchi and H.J. Schultz, Phys. Rev. {\bf
B61}, 13609 (2000); C.J. Halboth and W. Metzner, Phys. Rev. {\bf
B61}, 7364 (2000); C. Honerkamp et al, Phys. Rev. {\bf B63}, 035109
(2001).

\bibitem{8} P.W. Anderson, Science {\bf 235}, 1196 (1987).

\bibitem{ribeiro05}
T.~C. Ribeiro and X.~-G. Wen, Phys. Rev. Lett. {\bf 95}, 057001
(2005).




\bibitem{ribeiro06}
T.~C. Ribeiro and X.~-G. Wen, Phys. Rev. B {\bf 74}, 155113 (2006).



\bibitem{ferraz06} A. Ferraz, E. Kochetov, and M.Mierzejewski, Phys.
Rev. B {\bf 73}, 064516 (2006).



\bibitem{ferraz07}
A. Ferraz, E. Kochetov, and B. Uchoa, Phys. Rev. Lett. {\bf 98},
069701 (2007).

\bibitem{china}
After completion of this work we received an unpublished paper by
Qiang-Hua Wang, Fei Tan and Yuan Wan \cite{wang}, in which the
spinon-dopon NDO constraint (6) is enforced in a mean-field theory
by the requirement that the total number of the spinon-dopon
singlets should be equal to the averaged number of dopons. Within
that approach the authors have managed to explain  the so-called
waterfall anomaly in the nodal quasi particle dispersion observed in
hole doped cuprates.

\bibitem{wang}
Qiang-Hua Wang, Fei Tan and Yuan Wan, arXiv:cond-mat/0610491.


\bibitem{p} This projection operator plays a role similar
to that of a Gutzwiller projection in the conventional lattice
electron representation of the t-J model.



\bibitem{feng1} Shiping Feng, Tianxing Ma, and Jihong Qin, J. Phys.
Condens. Matter {\bf 16}, 343 (2004).






\bibitem{comment}
We mean a theory that incorporates the spin-dopon constraint at a
mean-field level. This should not be confused with Ribeiro and Wen
theory that neglects the constraint.



\bibitem{lnw} P.A. Lee, N. Nagaosa, and X.-G. Wen, Rev. Mod. Phys.
{\bf 78}, 17 (2006).

\bibitem{ichinose} I. Ichinose, T. Matsui, and M. Onoda, 
Phys. Rev. B  {\bf 64}, 104516 (2001).

\bibitem{zr}
F.S. Zhang and T.M. Rice, Phys. Rev. B  {\bf 37}, 3759 (1988).

\bibitem{lambda} Technically, this limit can be treated as follows.
We rotate the $d$-spinors to the $z$ axis, $d_i\to U_id_i,$ where
$U_i\tau_zU^{\dagger}_i=2\vec S_i\vec\tau$ \cite{wiegmann}. In this
way we arrive at the theory where the dopon propagator takes on the
form $(-\partial_t-\frac{3}{4}\lambda
+\mu-\frac{\lambda}{2}\tau_z)^{-1}$. Using this propagator
Abrikosov's fermionic diagramatic technique can be derived. This
technique involves infinitely large Lagrange multiplier that
modifies the fermionic propagator due to the extra term $\lambda
f^{\dagger}f$. This approach was succesfully used to describe the
Kondo effect in metals \cite{abrikosov} as well as thermodynamics of
the quantum Hisenberg model \cite{larkin} in the spinon
representation for the localized lattice spins, $\vec
S=f^{\dagger}\vec \tau f$, with the local constraint
$f^{\dagger}f=1$ imposed.

\bibitem{wiegmann} P.B. Wiegmann, Phys. Rev. Lett. {\bf 60}, 821
(1988).

\bibitem{abrikosov} A.A. Abrikosov, Physics {\bf 2}, 5 (1965).

\bibitem{larkin} V.G. Vaks, A.I. Larkin and S.A. Pikin, JETP {\bf
26}, 188 (1968).

\bibitem{tsunetsugu97}
H. Tsunetsugu, M. Sigrist, and K. Ueda, Rev. Mod. Phys. {\bf 69},
809 (1997).

\bibitem{referee} We would like to thank a Referee of our paper for
calling our attention to the issue on a competition between the RKKY
and Kondo effects for increasing doping.


\bibitem{senthil04}
T. Senthil, M. Vojta, and S. Sachdev, Phys. Rev. B {\bf 69}, 035111
(2004).



\bibitem{fulde06}
P. Fulde, P. Thalmeier, and G. Zwicknagl, Solid State Physics,
Volume 60 (Academic Press, 2006).

\bibitem{coleman} For review see P. Coleman, "Heavy Fermions:
electrons at the edge of magnetism" cond-mat/0612006.


\bibitem{paul07}
I. Paul, C. P\'epin, and M.~R. Norman, Phys. Rev. Lett. {\bf 98},
026402 (2007).



\bibitem{sidorov02}
V.~A. Sidorov \emph{et al.}, Phys. Rev. Lett. {\bf 89}, 157004
(2002).



\bibitem{bel04}
R. Bel \emph{et al.}, Phys. Rev. Lett. {\bf 92}, 217002 (2004).



\bibitem{nakajima07}
Y. Nakajima \emph{et al.}, Jour. Phys. Soc. Japan {\bf 76}, 024703
(2007).

\bibitem{mathur98}
N.~D. Mathur \emph{et al.}, Nature {\bf 394}, 39 (1998).



\bibitem{moriya03}
T. Moriya and K. Ueda, Rep. Prog. Phys. {\bf 66}, 1299 (2003).



\bibitem{wen HF} Interestingly, the similar
conclusion has been reached in \cite{ribeiro06} though through a
different line of arguments. In particular, the authors of this
paper make the following statement: " ...the generalized-tJ model
"doped carrier" formulation resembles that of heavy-fermion models:
dopons and lattice spins in the "doped carrier" framework correspond
to conduction electrons and to the spins of $f$-electrons,
respectively, in heavy-fermion systems. The main difference is that,
at low dopings, the tt't''J model spin-spin interaction is larger
than the dopon Fermi energy, while in heavy-fermion models the
spin-spin interaction between $f$-electrons is much smaller than the
Fermi energy of conduction electrons. As the doping concentracion
increases the dopon Fermi energy approaches, and can even overcome,
the interaction energy between lattice spins. In that case our
approach to the tt't''J model becomes qualitatively similar to
heavy-fermion models." Note that the NDO constraint discussed in the
present paper enables us to make this remark explicit.



























\bibitem{cobalt} In case the optimal doping is already large enough
(i.e., for the cobaltates), one should use the full spin-dopon
representation of the Heisenberg term as given by Eq. (\ref{tj5})
instead of the simplified one used in (\ref{eq:Hf}). In this case a
direct dopon-dopon interaction becomes relevant which may result in
the RVB-type dopon-dopon superconducting pair state \cite{ferraz06}.


\end{thebibliography}
\end{document}